\input gr-qc
%
\newdimen\LCBtmp%
\def\Vec#1{\mathchoice%
   {\maketildevec{\textstyle}{#1}}%
   {\maketildevec{\displaystyle}{#1}}%
   {\maketildevec{\scriptstyle}{#1}}%
   {\maketildevec{\scriptscriptstyle}{#1}}}%
\def\maketildevec#1#2{%
   \vtop{\baselineskip=0pt\lineskip=0pt%
   \setbox1=\hbox{$\seventeenpoint#1\char`\~$}%
   \LCBtmp=1.2\ht1%
   \setbox0=\hbox{\lower\LCBtmp\hbox{\box1}}\ht0=0pt\dp0=0pt%
   \ialign{\hfil ##\hfil\crcr\hbox{$#1#2$}\crcr\box0\crcr}}}%
%
%
\def\Tilde#1{\mathchoice%
   {\maketilde{\textstyle}{#1}}%
   {\maketilde{\displaystyle}{#1}}%
   {\maketilde{\scriptstyle}{#1}}%
   {\maketilde{\scriptscriptstyle}{#1}}}%
\def\maketilde#1#2{%
   \vbox{\baselineskip=0pt\lineskip=0pt%
   \setbox1=\hbox{$\seventeenpoint#1\char`\~$}%
   \LCBtmp=0.6\ht1%
   \setbox0=\hbox{\lower\LCBtmp\hbox{\box1}}\ht0=0pt\dp0=0pt%
   \ialign{\hfil##\hfil\crcr\box0\crcr\hbox{$#1#2$}\crcr}}}%

\advance\baselineskip+1.2pt
%
%
\def\eps{\epsilon}
\def\Oeps(#1){{\cal O}(\eps^{#1})}
\def\m{{\hbox{\tt-}}}
\def\p{{\hbox{\tt+}}}
\def\Bar{\overline}
%
%
\def\Displaylines#1{\halign{\hbox to\displaywidth{%
$\hfil\displaystyle##\hfil $}&\llap{##}\crcr #1\crcr}}
%
%
\title{%
{An ADM 3+1 formulation for}\cr
{Smooth Lattice General Relativity}\cr}

\address{%
\cr
{\rm Leo Brewin}\cr
\cr
{\sl Department of Mathematics}\cr
{\sl Monash University}\cr
{\sl Clayton, Vic. 3168}\cr
{\sl Australia}\cr
}

\beginabstract

A new hybrid scheme for numerical relativity will be presented. The scheme will
employ a 3-dimensional spacelike lattice to record the 3-metric while using the
standard 3+1 ADM equations to evolve the lattice. Each time step will involve
three basic steps. First, the coordinate quantities such as the Riemann and
extrinsic curvatures are extracted from the lattice. Second, the 3+1 ADM
equations are used to evolve the coordinate data, and finally, the coordinate
data is used to update the scalar data on the lattice (such as the leg lengths).
The scheme will be presented only for the case of vacuum spacetime though there
is no reason why it could not be extended to non-vacuum spacetimes. The scheme
allows any choice for the lapse function and shift vectors. An example for the
Kasner $T^3$ cosmology will be presented and it will be shown that the method
has, for this simple example, zero discretisation error.

\endabstract

\beginsection{Introduction}

In a recent paper \Ref(1) a new numerical scheme was developed and successfully
applied in the construction of time symmetric initial data for a Schwarzschild
black hole. The method was based on two key ideas, one to use short geodesic
segments as the legs of the lattice and two, to employ Riemann normal
coordinates local to each vertex.

Our primary aim in this paper is to present a hybrid scheme in which a
3-dimensional spacelike lattice can be evolved using the standard ADM vacuum 3+1
equations.

The general scheme envisaged in \Ref(1) can be summarised as follows. The lattice
is a network of vertices and legs and is fully specified by the connectivity
matrix and the leg lengths. To each vertex there is an associated computational
cell defined by some local sub-set of the lattice (eg.\ the legs and vertices of
the tetrahedra attached to this vertex, though larger subsets could be
considered). Riemann normal coordinates are employed in each computational cell
leading to a metric of the form
$$
g_{\mu\nu}(x) = g_{\mu\nu} - {1\over3}R_{\mu\alpha\nu\beta}\> x^\alpha x^\beta
                + \Oeps(3)
\eqno\eqndef{RNCmetric}
$$
where $g_{\mu\nu}$ is chosen to be ${\rm diag(1,1,1)}$ and $\eps$ is a typical
length scale for the computational cell. The crucial assertion that the legs are
geodesic segments of this metric leads directly to the following equation for the
leg length $L_{ij}$ between vertices $(i)$ and $(j)$
$$
L^2_{ij} = g_{\mu\nu}\Delta x^\mu_{ij}\Delta x^\nu_{ij}
          - {1\over3} R_{\mu\alpha\nu\beta}\> x^\mu_i x^\nu_i
                                              x^\alpha_j x^\beta_j
          + \Oeps(5)
\eqno\eqndef{LegsEqtn}
$$
where $\Delta x^\mu_{ij} = x^\mu_j - x^\mu_i$. For each computational cell there
would be a small set of such equations, which in \Ref(1) where referred to
as the smooth lattice equations. These could, in principle, be solved as
a coupled set of equations for (estimates of) the Riemann tensor and any coordinates
not fixed by gauge freedoms. The computations for any one cell are fully
decoupled from those of any other cell. Thus this is a local problem and should
be amenable to Newton-Raphson methods. It is at this point that the design of
the lattice becomes crucial. It must be such that the equations in each cell
yield unambiguous estimates for the Riemann tensor. If the system is singular
then there will exist continuous families of solutions for the curvatures. This
indicates that the lattice lacks sufficient information to tie down a unique
estimate (or at worst a finite discrete set of estimates) for the curvatures.
The solution then would be to add extra information to the lattice such as extra
leg lengths. The lattice chosen in \Ref(1) did not suffer from this problem.

However, subsequent experiments \Ref(2) have shown that many seemingly suitable
lattices (such as in Fig \figrfr{DiamondCell}) are singular. Following these
experiments a slight variation to the above scheme has proved to be very
successful. It entails the addition of extra information in the form of the
polar directions of the legs attached to the central vertex of the cell. Since
the connection vanishes at the origin (by definition of the RNC) this is
equivalent to specifying the spherical polar angles of the vertices at the ends
of these legs. This not only supplies the missing information but it also fully
decouples the set of smooth lattice equations.

The modified scheme works as follows. Each computational cell will, for the
remainder of this paper, be assumed to consist of the central vertex and the
tetrahedra attached to this vertex. For each cell there will be one central
vertex, for which we will always use the label 0, and a set of vertices on the
surface of the cell. To each surface vertex $(i)$ we assign the polar
coordinates $(\theta_i,\phi_i)$. With this, the equations (\eqnrfr{LegsEqtn})
for the radial legs are satisfied by
$$
\eqalign{%
x_i &= L_{oi}\cos\phi_i\sin\theta_i\cr
y_i &= L_{oi}\sin\phi_i\sin\theta_i\cr
z_i &= L_{oi}\cos\theta_i\cr}
\eqno\eqndef{PolartoRNC}
$$
where $L_{oi}$ is the length of the radial leg from the origin of the cell to
the surface vertex $(i)$. The $(x_i,y_i,z_i)$ are the Riemann normal coordinates
for vertex $(i)$. Next, the curvatures can be computed by solving, by a least
squares method, the leg length equations (\eqnrfr{LegsEqtn}) applied to the legs
on the surface of the computational cell. Least squares, or some other method,
will be necessary since these equations are almost certainly overdetermined.

In this scheme the basic data given on the lattice are the $L_{ij}, \theta_i$
and $\phi_i$ from which we compute, by the above equations, the $x^\mu_i$ and
the $R_{\mu\nu\alpha\beta}$.

Note that for each leg on the surface of the cell there is an associated angle
subtended at the origin of the cell. In terms of the spherical polar
coordinates for the vertices $(i)$ and $(j)$ of the leg $(ij)$, this angle,
$\alpha_{ij}$, can be computed from
$$
\cos\alpha_{ij} = \cos\theta_i\cos\theta_j 
                + \cos(\phi_i - \phi_j)\sin\theta_i\sin\theta_j
\eqno\eqndef{AlphaMain}
$$
and in terms of the Riemann normal coordinates from
$$
\cos\alpha_{ij} = g_{\mu\nu} {\Delta x^\mu_{oi}\Delta x^\nu_{oj}\over
                              L_{oi}L_{oj}}
\eqno\eqndef{AlphaAlt}
$$
Both of these equations will be used later in this paper.

Note that in the construction of our Riemann normal coordinates we have taken
advantage of some coordinate freedoms. For example we have used translational
freedoms to tie the origin of the RNC to the central vertex. We have also chosen
our coordinate axes to be orthogonal at the origin (hence $g_{\mu\nu} = {\rm
diag}(1,1,1)$). However, there are some residual gauge freedoms, namely the
three degrees of freedom to rotate the coordinate axes about the origin. This
freedom can be accounted for by specifying three of the polar angles $\theta_i$
and $\phi_i$, such as $\theta_1,\phi_1$ and $\theta_2$.

\beginsection{Smooth lattice kinematics}

In the familiar 3+1 formulation of general relativity a typical spacetime can be
represented in two equivalent ways, one as an evolving 3-metric on 3-dimensional
manifold and two, as sequence of distinct 3-dimensional hypersurfaces of the
spacetime. In the later picture, each hypersurface records one instant in the
evolution of the 3-metric. The information that one must specify to fully define the 4-metric on
the spacetime are the 3-metric, the lapse function and the shift vectors on each
hypersurface. In this section we shall adapt this picture to one in which each
hypersurface is a smooth 3-dimensional lattice.

The general idea will be to allow the basic data for each lattice, namely the
$L_{ij}, \theta_i$ and $\phi_i$, to be functions of time. On each hypersurface
we will continue to employ Riemann normal coordinates and the above equations for
computing the curvature components. Since the lattice data are now presumed to
be functions of time so too must the coordinates $x^\mu_i$ and the curvature
components $R_{\mu\nu\alpha\beta}$. We will also impose our previously stated
gauge conditions on each hypersurface, in particular that $g_{\mu\nu}={\rm
diag}(1,1,1)$ and that the origin of the RNC remains forever tied to the central
vertex of the cell. One could choose a dynamical set of gauge conditions but
to do so at this stage would be an un-necessary complication.

It is customary to use Latin indices to denote spatial components (eg.\ $g_{ij}$
for the 3-metric) and Greek indices for spacetime quantities (eg.\ 
$g_{\mu\nu}$ for the 4-metric). As we would like to use Latin indices to
denote simplices of the lattice we shall choose to use Greek indices for
spatial components. Thus in the following $g_{\mu\nu}, R_{\mu\nu\alpha\beta}$
etc. will denote 3-tensors that live in each 3-dimensional hypersurface. 

We will be making frequent reference to the current and future hypersurfaces
and the various structures that define and link these hypersurfaces. We will use
$\Sigma_0$ to represent the current hypersurface and $\Sigma_{\delta t}$ to
represent the hypersurface just slightly to the future of $\Sigma_0$.  The time
coordinate $t$ is constant on each hypersurface with $t=t_0$ on $\Sigma_0$ and
$t=t_0 + \delta t$ on $\Sigma_{\delta t}$. The lapse function will be denoted
by $N$ and the shift vector by $N^\mu$. Both the lapse function and shift
vectors will be defined on the vertices of the lattice. Note that even though we
are using Riemann normal coordinates on each hypersurface the 4-dimensional
coordinates $(t,x^\mu)$ will not, in general, be in Riemann normal form.

Consider any 3-dimensional smooth lattice and focus attention on one vertex.
Through this vertex one can construct three well defined curves -- the timelike
worldline of the vertex, the timelike geodesic tangent to the normal vector at
the vertex and the timelike worldline of the observer with constant spatial
coordinates. It is important to realise that these may be three distinct curves.
Thus upon extending these curves from their common origin in $\Sigma_0$ to
$\Sigma_{\delta t}$ we can expect to obtain three new points in
$\Sigma_{\delta t}$. This situation is depicted in Fig (\figrfr{DriftVec}). In
the usual 3+1 picture only two of these curves, the normal geodesic and the
observers worldline, enter into the discussion. In the smooth lattice we must
also consider the freedom for vertices to drift relative to these two curves. For
this reason we will introduce a new vector, the drift vector $\gamma^\mu$.

From Fig (\figrfr{DriftVec}) we can clearly see that the shift and drift vectors
at vertex $(i)$ are related by
$$
{dx^\mu_i\over dt} = -\gamma^\mu_i + N^\mu_i
\eqno\eqndef{DriftVecA}
$$
where $x^\mu_i(t)$ are the time dependent coordinates of vertex $(i)$. The
$dx^\mu_i/dt$ can be easily computed from (\eqnrfr{PolartoRNC}). Note that in
our chosen gauge $x^\mu_i=dx^\mu_i/dt=0$ at the origin of the RNC.

One now has a choice as to which of the drift or shift vectors should be freely
specified on the lattice. Since the drift vector has a strong geometric appeal
we shall choose to freely specify it and to use the above equation
(\eqnrfr{DriftVecA}) to compute the shift vector. A convenient choice is to set
the drift vector to be zero at the origin (ie.\ the vertex at the origin of the
RNC in $\Sigma_0$ is evolved along the timelike normal). Though, of course, there
may be occasions where such a choice is not appropriate.

We will now introduce a second lattice which will help us later on in making the
transition to the standard ADM 3+1 equations. Consider one hypersurface and its
associated lattice. Consider now a second lattice coincident with the original
lattice. We will refer to these as the primary and shadow lattices respectively.
The shadow lattice will be chosen so such that its vertices are evolved along
the normals to the hypersurface (ie.\ as if the drift vectors were set to zero
everywhere). We do this because the ADM 3+1 equations take on a particularly
simple form when expressed as time derivatives along the normals. The shadow
lattice will only be employed for one time step. Upon completion of the time
step the current shadow lattice will be discarded and a new shadow lattice
created coinciding with the updated primary lattice. Each of these shadow
lattices are introduced only as an aid in the exposition of our algorithm --
they need never be created in any computer program.

Our current task is to see how we should specify the data on the shadow lattice
so that it evolves along the normals.

On the primary lattice we have various quantities such as $L_{ij}, \theta_i,
\phi_i$ and  $x^\mu_i$ all of which we assume to be functions of time $t$. Their
counterparts on the shadow lattice will be denoted by the addition of a dash
(ie.\ $L'_{ij}, x'^\mu_i$ etc.). The dash on the $x'^\mu_i$ should not be
confused with coordinate transformation -- the $x'^\mu_i$ are just the $x^\mu$
coordinates of vertex $(i')$. On the initial hypersurface $\Sigma_0$ all of
the corresponding dashed and un-dashed quantities are equal (ie.\ $L'_{ij}=L_{ij},
x'^\mu_i=x^\mu_i$) though they may drift apart between successive hypersurfaces.

By inspection of Fig (\figrfr{DriftVec}) one can easily verify that
$$
{dx'^\mu_i\over dt} = {dx^\mu_i\over dt} - \gamma^\mu_i
\eqno\eqndef{DriftVecB}
$$

Consider now equation (\eqnrfr{LegsEqtn}) applied to both lattices,
$$
\eqalign{%
L^2_{ij}(t) &= g_{\mu\nu}\Delta x^\mu_{ij}\Delta x^\nu_{ij}
             - {1\over3} R_{\mu\alpha\nu\beta}\> x^\mu_i x^\nu_i
                                              x^\alpha_j x^\beta_j\cr
{L'}^2_{ij}(t) &= g_{\mu\nu}\Delta x'^\mu_{ij}\Delta x'^\nu_{ij}
             - {1\over3} R_{\mu\alpha\nu\beta}\> x'^\mu_i x'^\nu_i
                                              x'^\alpha_j x'^\beta_j\cr
}
$$
All of the terms on the right hand side should be considered to be functions of
time (except the $g_{\mu\nu}$ which we choose to be $\rm diag(1,1,1)$).
Differentiating with respect to $t$ and using the above equation for
$dx'^\mu_i/dt$ we obtain on $\Sigma_0$
$$
{dL'_{ij}\over dt} = {dL_{ij}\over dt} 
        - {1\over  L_{ij}} g_{\mu\nu}\Delta\gamma^\mu_{ij}\Delta x^\nu_{ij}
        + {1\over 3L_{ij}} R_{\mu\alpha\nu\beta}\>( \gamma^\mu_i x^\nu_i
                                                    x^\alpha_j x^\beta_j
                                                  + x^\mu_i x^\nu_i
                                                    \gamma^\alpha_j x^\beta_j )
\eqno\eqndef{ShadowLeg}
$$
where $\Delta\gamma^\mu_{ij} = \gamma^\mu_j - \gamma^\mu_i$.

A similar argument can be applied to the angles $\alpha_{ij}$. Starting with
$$
\eqalignno{%
\cos\alpha_{ij}(t) &= g_{\mu\nu} {\Delta x^\mu_{oi}\Delta x^\nu_{oj}\over
                              L_{oi}L_{oj}}
&\eqndef{ShadowAngA}\cr
\cos\alpha'_{ij}(t) &= g_{\mu\nu} {\Delta x'^\mu_{oi}\Delta x'^\nu_{oj}\over
                              L'_{oi}L'_{oj}}
&\eqndef{ShadowAngB}\cr
}
$$
we obtain, by direct differentiation, that on $\Sigma_0$
$$
\eqalign{%
  {d\alpha'_{ij}\over dt}
 = {d\alpha_{ij}\over dt}
&+ {\cos\alpha_{ij}\over\sin\alpha_{ij}}
     \left( {1\over L_{oi}}{d\over dt}(L'_{oi}-L_{oi})
           +{1\over L_{oj}}{d\over dt}(L'_{oj}-L_{oj}) \right)\cr
&+ {g_{\mu\nu}\over L_{oi}L_{oj}\sin\alpha_{ij}}
    \left( \Delta\gamma^\mu_{oi}\Delta x^\nu_{oj}
          +\Delta\gamma^\mu_{oj}\Delta x^\nu_{oi}\right)\cr}
$$
This can be simplified by using the above equation for $dL'/dt$ and noting that
in our gauge $x^\mu_o = 0$ on $\Sigma_0$. The result is that on $\Sigma_0$
$$
\eqalign{%
   {d\alpha'_{ij}\over dt}
 = {d\alpha_{ij}\over dt}
+ {g_{\mu\nu}\over L_{oi}L_{oj}\sin\alpha_{ij}}
   \Big(
&\phantom{+}
    \left( \Delta\gamma^\mu_{oi}\Delta x^\nu_{oj}
          +\Delta\gamma^\mu_{oj}\Delta x^\nu_{oi}
    \right)\cr
&-  \cos\alpha_{ij}
    \left( {L_{oj}\over L_{oi}}\Delta\gamma^\mu_{oi}\Delta x^\nu_{oi}
          +{L_{oi}\over L_{oj}}\Delta\gamma^\mu_{oj}\Delta x^\nu_{oj}
    \right)
   \Big)\cr
}
\eqno\eqndef{ShadowAng}
$$

The $dL'_{ij}/dt$ and the $d\alpha'_{ij}/dt$ measure rates of change along the
normals to $\Sigma_0$. They should therefore be expressible in terms of the
extrinsic curvatures $K'_{\mu\nu}$. 

Consider the shadow lattice and the coordinates $x^\mu$. Let us introduce
a second coordinate system $x''^\mu$ such that the $x''^\mu$ are constant along
each worldline of the vertices of the shadow lattice. This clearly
establishes a time dependent coordinate transformation between $x^\mu$ and
$x''^\mu$. Note that on $\Sigma_0$ both coordinate systems are identical. Note
also that on subsequent hypersurfaces the $x''^\mu$ need not be in Riemann
normal coordinate form.

Since the worldlines of the shadow lattice are tangent to the normals to
$\Sigma_0$ we must have
$$
{dg''_{\mu\nu}\over dt} = - 2 N K''_{\mu\nu}
$$

In the $x''^\mu$ coordinates the $L'_{ij}(t)$ and $\alpha'_{ij}(t)$ may be
estimated from
$$
\eqalignno{%
{L'}^2_{ij}(t) &=
     g''_{\mu\nu}(t) \Delta x''^\mu_{ij}\Delta x''^\nu_{ij} + \Oeps(4)
&\eqndef{LsqGddash}\cr
L'_{oi}L'_{oj}\cos\alpha'_{ij}(t) &= 
     g''_{\mu\nu}(t) \Delta x''^\mu_{oi}\Delta x''^\nu_{oj} + \Oeps(4)
&\eqndef{AlphaGddash}\cr
}
$$
where $g''_{\mu\nu}$ is evaluated on the geodesic tangent to the normal to
$\Sigma_0$ at the origin of the RNC. Since the $x''^\mu$ need not be in
Riemann normal coordinate form we can expect these estimates to be
less accurate than their Riemann normal coordinate counterparts (eg.\ equations
(\eqnrfr{LegsEqtn}) and (\eqnrfr{AlphaAlt})). This is the price we pay for
not using Riemann normal coordinates. Increased accuracy could be obtained for
the $L'^2_{ij}(t)$ by evaluating the right hand side at the centre of the legs.
This then forces us to explicitly account for variations of $K_{\mu\nu}$
throughout the cell. It is not clear how one might make a similar `centred'
approximation for the angles $\alpha'_{ij}(t)$. Rather than trying to resolve
these issues in this paper we shall settle on the simple estimates given
above.

Differentiating the previous two equations with respect to time and using the
above equation for $K''_{\mu\nu}$ we obtain, on $\Sigma_0$,
$$
\eqalignno{%
{L'}_{ij}{dL'_{ij}\over dt} &= - (N K_{\mu\nu})_o
                                 \Delta x^\mu_{ij}
                                 \Delta x^\nu_{ij}
&\eqndef{KmunuLegs}\cr
{d\over dt}\left(L'_{oi}L'_{oj}\cos\alpha'_{ij}\right)
   &= - 2 (N K_{\mu\nu})_o \Delta x^\mu_{oi}\Delta x^\nu_{oj}
&\eqndef{KmunuAngles}\cr
}
$$
This system of equations for the six $K_{\mu\nu}$ at the origin of the RNC will,
for almost all lattices, be highly overdetermined (there being many more
$L_{ij}$ and $\alpha_{ij}$ than the six $K_{\mu\nu}$). This problem can be
overcome in one of two ways. We could reject all but six of the above equations.
Or we could use all of the equations in a least squares estimation of the six
$K_{\mu\nu}$. We will assume in the following that the least squares method has
been used. In fact we will find that this issue of overdeterminism will crop up
on a number of occasions in this paper. In all cases we will resort to a least
squares solution as it is systematic and easy to implement in a computer program.

At this point we are in a position to calculate both the extrinsic and
intrinsic curvatures tensors, at the origin of the RNC, given just the basic
lattice data, namely the $L_{ij}, \theta_i, \phi_i$, their time derivatives and
for any choice of chosen lapse function and shift and drift vectors. However we
have not discussed how $K_{\mu\nu\vert\alpha}$ and $N_{\vert\mu\nu}$ may be
calculated (both of which will be needed in the next section).

Consider first the calculation of $N_{\vert\mu\nu}$. Since $N$ is a scalar
function we have $N_{\vert\mu} = \partial N/\partial x^\mu$ which we
can approximate at the centre of the leg $(ij)$ by
$$
\left(N_{\vert\mu}\right)_{ij} = {N_i - N_j\over L_{ij}}\Delta x^\mu_{ij}
\eqno\eqndef{LapseDerivA}
$$
Such an expression can be formed at the centre of each leg of the computational
cell. From this data we can form a linear approximation
${\Tilde N}_{\vert\mu}(x^\alpha)$ to $N_{\vert\mu}(x^\alpha)$
$$
{\Tilde N}_{\vert\mu}(x^\alpha) = {\Tilde N}_{\vert\mu} 
                                + {\Tilde N}_{\vert\mu\nu} x^\nu
\eqno\eqndef{LapseDerivB}
$$
where the constants ${\Tilde N}_{\vert\mu}$ and
${\Tilde N}_{\vert\mu\nu} = {\Tilde N}_{\vert\nu\mu}$ are obtained by a least
squares fit. We take the ${\Tilde N}_{\vert\mu\nu}$ as our estimates
of $N_{\vert\mu\nu}$ at the origin of the RNC. Similarly the
${\Tilde N}_{\vert\mu}$ could be used as an approximation to $N_{\vert\mu}$ at
the origin of the RNC (though we will not need to do so). The appropriate least
squares sum is chosen to be
$$
S({\Tilde N}_{\vert\mu},{\Tilde N}_{\vert\mu\nu}) =
\sum_\mu\sum_{ij}\left( \left(N_{\vert\mu}\right)_{ij} 
                       - {\Tilde N}_{\vert\mu}
                       - {1\over2}{\Tilde N}_{\vert\mu\nu}{(x^\nu_i + x^\nu_j)}
                 \right)^2
$$
Note that the symmetry condition
${\Tilde N}_{\vert\mu\nu} = {\Tilde N}_{\vert\nu\mu}$ must be explicitly imposed
in this sum. 

Do we have enough data to compute the nine numbers ${\Tilde N}_{\vert\mu}$ and
${\Tilde N}_{\vert\mu\nu}$? The answer is yes -- each leg gives us three samples
for $N_{\vert\mu}$ and each computational cell is certain to have more than
three legs.

The computation of the $K_{\mu\nu\vert\alpha}$ can be performed in a similar
manner. The first thing to do is to compute the $K_{\mu\nu}$ at the origin of
each RNC for each computational cell. Now consider one specific computational
cell. The vertices on the surface of the cell will themselves be the origins of
neighbouring cells. Each of these cells carries their own estimates of the
$K_{\mu\nu}$ in their own coordinates. As there is a well defined coordinate
transformation between these neighbouring frames it is possible to obtain
estimates of the $K_{\mu\nu}$ on each of the vertices in our chosen cell and in
the coordinates of that chosen cell (as indicated in Fig (\figrfr{KmunuFitA})).
We can then form a linear approximation to each of the six $K_{\mu\nu}$ in the
chosen cell. Thus we put
$$
{\Tilde K}_{\mu\nu}(x^\alpha) = {\Tilde K}_{\mu\nu} 
                              + {\Tilde K}_{\mu\nu\vert\alpha}x^\alpha
$$
with the constants ${\Tilde K}_{\mu\nu}$ and ${\Tilde K}_{\mu\nu\vert\alpha}$
obtained by another application of the least squares method. Finally we take the
${\Tilde K}_{\mu\nu\vert\alpha}$ as our estimate of $K_{\mu\nu\vert\alpha}$ at
the origin of our chosen cell. Note that in this case we would solve six separate
least squares problems, one for each of the six ${\Tilde K}_{\mu\nu}(x^\alpha)$.
In each case we need to compute just four numbers, one ${\Tilde K}_{\mu\nu}$ and
three ${\Tilde K}_{\mu\nu\vert\alpha}$. Once again we see that the least squares 
method is appropriate.

\beginsection{Smooth lattice dynamics}

In the standard ADM 3+1 formulation of vacuum General Relativity there are
four constraint equations
$$
\openup10pt\Displaylines{%
0 = R + K^2 - K_{\mu\nu}K^{\mu\nu}&\eqndef{HamltnConst}\cr
0 = K_{\vert\mu} - K^\nu_{\mu\vert\nu}&\eqndef{MomtmConst}\cr
}
$$
and six evolution equations
$$
\openup10pt\displaylines{%
{\partial g_{\mu\nu}\over \partial t} =
 - 2 N K_{\mu\nu} + \beta_{\mu\vert\nu} + \beta_{\nu\vert\mu}\cr
{\partial K_{\mu\nu}\over \partial t} =
 - N_{\vert\mu\nu} + N \left( R_{\mu\nu} +  K K_{\mu\nu}
                                         - 2K_{\mu\alpha}K^\alpha_\nu \right)
 + N^\alpha K_{\mu\nu\vert\alpha} + K_{\mu\alpha} N^\alpha_{\vert\nu}
                                  + K_{\nu\alpha} N^\alpha_{\vert\mu}\cr
}
$$
where $K=K^\alpha_\alpha, R_{\mu\nu}=g^{\alpha\beta}R_{\alpha\mu\beta\nu}$ and
$R=g^{\mu\nu}R_{\mu\nu}$.

Let us concentrate, for the moment, on the evolution equations.
When cast as evolution equations along the timelike normals these equations take
on a particularly simple form, namely,
$$
\openup10pt\eqalignno{%
{dg_{\mu\nu}\over dt} &= - 2 N K_{\mu\nu}
&\eqndef{DgDt}\cr
{dK_{\mu\nu}\over dt} &=
 - N_{\vert\mu\nu} + N \left( R_{\mu\nu} + K K_{\mu\nu}
                                       -2K_{\mu\alpha}K^\alpha_\nu \right)
&\eqndef{DkDt}\cr
}
$$
The easiest road to these equations is to simply put the shift vector to zero
in the original equations. However these equations can also be easily derived by
direct calculation.

The set of equations from (\eqnrfr{LegsEqtn}) through to (\eqnrfr{DkDt}) can be
viewed as a large set of coupled ordinary differential equations controlling
the future evolution of the lattice. These equations will need to be solved by
some numerical integration scheme to generate the lattice at successive time
steps. For pure simplicity we shall demonstrate one time step using the naive
Euler method, namely, that  each quantity is updated according to
$x\leftarrow x + {\dot x}\delta t$. Note that we do not advocate the use of
this naive integration scheme in any realistic implementation (suitable schemes
would include predictor-corrector, leapfrog or Runge-Kutta methods).

We will start with the lattice data on $\Sigma_0$, namely, all of the
$L_{ij}, \theta_i, \phi_i, dL_{ij}/dt, d\theta_i/dt$ and $d\phi_i/dt$. We will
also assume that well defined rules are given for choosing the $N, N^\mu$
and $\gamma^\mu$ on each vertex of the lattice for all time. We will finish
with this data fully evolved to the next hypersurface $\Sigma_{\delta t}$.

Here is our proposed algorithm.

\advance\leftskip+1.0cm

\def\A#1{\hbox to 0.8cm{\hfil#1}}

\item{1.} For each computational cell

\advance\leftskip+1.0cm
\item{\A{1.1}} Use equations (\eqnrfr{LegsEqtn},\eqnrfr{PolartoRNC})
               to compute the $x^\mu_i, dx^\mu_i/dt$
               and the $R_{\mu\nu\alpha\beta}$. 
\item{\A{1.2}} Use equations (\eqnrfr{AlphaMain}) and its time derivatives
               to compute the $\alpha_{ij}$ and $d\alpha_{ij}/dt$. 
\item{\A{1.3}} Use equations (\eqnrfr{ShadowLeg},\eqnrfr{ShadowAng})
               to compute the $dL'_{ij}/dt$ and $d\alpha'_{ij}/dt$
\item{\A{1.4}} Use equations (\eqnrfr{KmunuLegs},\eqnrfr{KmunuAngles})
               to compute the $K_{\mu\nu}$
\item{\A{1.5}} Use equations (\eqnrfr{LapseDerivA},\eqnrfr{LapseDerivB})
               to compute $N_{\vert\mu\nu}$.
\item{\A{1.6}} Update $L_{ij}$ using $dL_{ij}/dt$.
\item{\A{1.7}} Update $x^\mu_i$ using $dx^\mu_i/dt$.
\item{\A{1.8}} Update $\theta_i,\phi_i$ using $d\theta_i/dt,d\phi_i/dt$.
\item{\A{1.9}} Update $K_{\mu\nu}$ using equations (\eqnrfr{DkDt}).

\advance\leftskip-1.0cm

\item{2.} Interpolate $K_{\mu\nu}$ back to the primary lattice.
\item{3.} For each computational cell

\advance\leftskip+1.0cm
\item{\A{3.1}} Update $dL'_{ij}/dt$ using equations (\eqnrfr{KmunuLegs}).
\item{\A{3.2}} Update $dL_{ij}/dt$ using equations (\eqnrfr{ShadowLeg}).
\item{\A{3.3}} Update $d\alpha'_{ij}/dt$ using equations (\eqnrfr{KmunuAngles}).
\item{\A{3.4}} Update $d\alpha_{ij}/dt$ using equations (\eqnrfr{ShadowAng}).
\item{\A{3.5}} Update $d\theta_i/dt,d\phi_i/dt$ using
               the time derivative of equations (\eqnrfr{AlphaMain}).

\advance\leftskip-2.0cm

The are a number of points that need to be made. As each leg will appear in one
or more computational cells there is an ambiguity in how $L_{ij}$ and
$dL_{ij}/dt$ should be updated. For example, in the cubic lattice of Fig
(\figrfr{CubicLattice}), each leg is contained in exactly two different
computational cells. In this case the $L_{ij}$ and $dL_{ij}/dt$ could be updated
as the average from both cells. Other options are possible and should be
explored by direct numerical experimentation. Notice that there is no such
ambiguity in updating the angles $\alpha_{ij}$ since these are defined at the
origin of each computational cell.

The calculations required in step 2 are non-trivial and arise because the shadow
lattice will drift from the primary lattice (see Fig (\figrfr{KmunuFitB})). As
with the computation of the $K_{\mu\nu\vert\alpha}$ we would need to use
coordinate transformations between neighbouring cells and a least squares linear
approximation to interpolate the $K_{\mu\nu}$ from the shadow lattice back to
the primary lattice. If one chooses the drift vector to be zero at the origin of
each cell then this step is not needed and steps 1 and 3 can be merged as one
larger step. This is a considerable computational advantage. But does it
restrict the class of spacetimes that can be built by this method? Setting the
drift vector to zero everywhere on a lattice is much the same as setting the
shift vector to zero everywhere on a finite difference grid. We know that in the
later case the lapse function may need to be carefully chosen so as to avoid the
numerical grid collapsing on itself (eg.\ as occurs for $N=1, N^\mu=0$ in a
Schwarzschild spacetime). The same should apply to a lattice -- the lapse will
need to be carefully chosen to ensure that the lattice does not collapse on
itself. There are also good arguments for using a shift vector to encourage the
grid points to maintain a reasonable structure (eg.\ that the density of grid
points does not drift too far from some preferred arrangement). It seems
reasonable to expect that many interesting spacetimes can be explored using a
zero drift vector and an appropriately chosen lapse function. Despite this,
there are known cases (in finite difference calculations) where a non-zero shift
vector is highly desirable. For example, in the apparent horizon boundary
condition used by Anninos et al \Ref(3) a non-zero shift vector is explicitly
used to ensure that grid points do not fall into a black hole and to retain a
good resolution of the metric in the vicinity of the apparent horizon. Such a
condition led to a dramatic improvement in the long term stability of the
numerical evolution. We can expect that a similar condition will be needed in
the smooth lattice approach (though this has yet to be tested).

Another important issue is to what extent does the extensive use of least
squares approximations (which are required in steps 1.1, 1.4, 1.5, 2, 3.5) effect the accuracy and
stability of the evolution? The least squares approximations are bound to
introduce some degree of smoothing in the estimates of $N_{\vert\mu\nu},
K_{\mu\nu}$ etc. Does this wash out unimportant numerical noise or important
short scale variations in the lattice data? Most likely both important and
unimportant information is lost. The question is does this effect the long term
stability of the evolution? Again this can be investigated by way of a number of
numerical experiments on a series of successively refined lattices.

The reason that we advocated the use of least squares was that it overcame
the problems of estimating various quantities from an overdetermined system of
equations. One could imagine a highly idealised case of an overdetermined system
which admits a well defined consistent solution. For
example four points in a plane constitute an overdetermined system for a
straight line. But properly chosen those four points will determine one straight
line. In all other cases a straight line approximation could be constructed by a
least squares method. Suppose now that we wished to impose some dynamics on this
toy problem. Let us suppose that we started with four points that lie on one straight
line (ie.\ a consistent yet overdetermined system). If we impose individual evolution
equations for each of the four points then those four points may no longer consistently
determine a unique straight line. If however we impose the evolution equations
on the parameters of the straight line then we can consistently evolve all four
points. This same observation applies to our evolution of the smooth lattice.
It would be wrong to search directly for evolution equations for each of the
leg lengths. In our scheme we impose the dynamics on the $K_{\mu\nu}$ which we
extract from an overdetermined system of equations on the lattice data. Our
hope is that if we start with a lattice for which we have a good least squares
fit for $K_{\mu\nu}$ etc. then the evolution will continue to yield a good fit.
Again this is speculative and must be tested by direct numerical computation.

The discussion so far has dealt entirely with the evolution of the lattice. Some
words are thus in order regarding the construction of initial data on the
lattice. We have already seen how all of the quantities needed for the constraint
equations can be extracted from the lattice. Thus we are able to evaluate the
right hand side of the constraint equations. Should this be non-zero then clearly
we do not have valid initial data and some corrections must be made. Following
the philosophy espoused in the previous paragraph the corrections to the lattice
data should be driven by corrections in the coordinate data. How this might be
achieved is far from clear. Can an approach similar to York's conformal method
be used? Part of that method is to propose a 3-metric of the form $g_{\mu\nu} =
\phi^4 {\Tilde g}_{\mu\nu}$ where ${\Tilde g}_{\mu\nu}$ is a given 3-metric
(commonly chosen to be flat) and $\phi$ is the conformal factor. With two
metrics one has a choice as to which should be expressed in Riemann normal form.
If we choose ${\Tilde g}$ then we can imagine also having a conformal lattice
with leg lengths ${\Tilde L}_{ij}$. The $L_{ij}$ of the physical lattice could
then be estimated as $L_{ij} = \phi_i\phi_j {\Tilde L}_{ij}$ (as suggested
by Piran and Williams \Ref(4)). It is unclear at present how York's transverse traceless tensor
can be represented on the conformal lattice. On the other hand if one chooses to
express the physical metric $g_{\mu\nu}$ in Riemann normal form then what form
does a flat ${\Tilde g}_{\mu\nu}$ take in these coordinates? We conclude that it
may not be a simple task to apply York's method to a smooth lattice.

Another important question concerns the order of the discretisation errors.
Previous calculations (see \Ref(1)) showed that the smooth lattice method
yielded $\Oeps(2)$ accurate estimates for both the metric and the Riemann
tensors. However, this is unlikely to be the case for generic lattices. The
formal truncation error in the Taylor series expansion of the metric, equation
(\eqnrfr{RNCmetric}), is $\Oeps(3)$ and thus we can expect the discretisation
error in the curvatures to be of $\Oeps(1)$. This will couple to the smooth
lattice equations leading to an expected error in the metric of $\Oeps(1)$. This
is less than optimal -- we would prefer an $\Oeps(2)$ error for generic
lattices. The challenge then is to find ways in which second order accurate
estimates for the curvatures on the lattice can be obtained. This also applies
to other term such as $N_{\vert\mu\nu}$ and $K_{\alpha\beta\vert\mu}$. We hope to
report on this in a later paper.

\beginsection{An example : The Kasner cosmology}

The Kasner metric
$$
ds^2 = - (dt')^2 + f^2_x(t')(dx')^2 + f^2_y(t')(dy')^2 + f^2_z(t')(dz')^2
$$
describes a homogeneous cosmology with a $T^3$ topology. This metric, for the
particular choice $f_x(t') = t^{p_x}, f_y(t') = t^{p_y}, f_z(t') = t^{p_z}$, is
a solution of the vacuum Einstein equations when
$$
1 = p_x + p_y + p_z = p^2_x + p^2_y + p^2_z
$$

A trivial transformation of the spatial coordinates
$$
\eqalignno{%
x &= f_x(t')(x'-x'_o)\cr
y &= f_y(t')(y'-y'_o)\cr
z &= f_z(t')(z'-z'_o)\cr}
$$
where $(x,y,z)'_o$ is any nominated point, leads to a 3-metric which, at the
chosen point, is in Riemann normal form with $g_{ij} = {\rm diag}(1,1,1)$.
This fact and the clear simplicity of the metric suggests that the 3+1 smooth
lattice approach should yield accurate approximations to the Kasner metric.

In fact this problem is so simple that we will be able to show that there is no
discretisation error. Of course this fortuitous result can not be expected to
occur in other less symmetric spacetimes.

The 3-dimensional lattice is chosen to be a cubic lattice with the addition of 
two extra diagonal legs to each 2-dimensional face. The typical RNC cell will
then consist of the central vertex, its six nearest neighbours, the six legs
that join them and the twelve diagonal legs not attached to the central vertex
(see Figs (\figrfr{DiamondCell}--\figrfr{CubicLattice})). The data for each RNC
cell will be the 6 radial leg lengths $L_{oi}$, 12 diagonal leg lengths $L_{ij}$,
the 6 angles $\alpha_{ij}$ and their respective time derivatives,
$dL_{oi}/dt, dL_{ij}/dt$ and $d\alpha_{ij}/dt$.

Throughout the evolution we will impose the following gauge freedoms.
\begnarrow
\item{\Mymark} The lapse function has the value one everywhere,
$N_i = 1$.
\item{\Mymark} The shift vector is zero at the origin of each RNC,
$0= N^\mu_o$.
\item{\Mymark} The drift vector is zero everywhere,
$0=\gamma^\mu_i$.
\endnarrow

While on the initial $t=t_0$ hypersurface we will assume
\begnarrow
\item{\Mymark} The diagonal legs are constrained by
$L^2_{ij} = L^2_{oi} + L^2_{oj}$.
\item{\Mymark} The angles $(\theta_i,\phi_i)$ are constrained such that
$\alpha_{ij} = {\pi/2}$ and ${d\alpha_{ij}/dt} = 0$.
\endnarrow
The choice $\alpha_{ij}=\pi/2$ corresponds to placing the vertices on the
coordinate axes while $d\alpha_{ij}/dt = 0$ ties them to the axes (though they
are free to move subsequently off the axes). The constraint on the diagonals
ensures that the 3-lattice is initially flat, ie.\ $R_{\mu\nu\alpha\beta}=0$.

We will defer stating the homogeneity conditions until we have explicitly
calculated all of the $K_{\mu\nu}$.

Starting with equations (\eqnrfr{LegsEqtn}--\eqnrfr{DriftVecA}) we quickly
obtain
$$
\Displaylines{%
R_{\mu\nu\alpha\beta} = 0\cr
\noalign{\vskip10pt}
(x^\mu_{1^\p}) = (L_{1^\p},0,0)\hskip 2cm (x^\mu_{1^\m}) = (-L_{1^\m},0,0)\cr
(x^\mu_{2^\p}) = (0,L_{2^\p},0)\hskip 2cm (x^\mu_{2^\m}) = (0,-L_{2^\m},0)\cr
(x^\mu_{3^\p}) = (0,0,L_{3^\p})\hskip 2cm (x^\mu_{3^\m}) = (0,0,-L_{3^\m})\cr
\noalign{\vskip10pt}
(N^\mu_{1^\p}) = ({\dot L}_{1^\p},0,0)\hskip 2cm
(N^\mu_{1^\m}) = (-{\dot L}_{1^\m},0,0)\cr
(N^\mu_{2^\p}) = (0,{\dot L}_{2^\p},0)\hskip 2cm
(N^\mu_{2^\m}) = (0,-{\dot L}_{2^\m},0)\cr
(N^\mu_{3^\p}) = (0,0,{\dot L}_{3^\p})\hskip 2cm
(N^\mu_{3^\m}) = (0,0,-{\dot L}_{3^\m})\cr}
$$
where a dot denotes $d/dt$.

Since $\gamma^\mu_i=0$ for $t\ge t_0$ while $d\alpha_{ij}/dt = 0$ on $t=t_0$ we
find from (\eqnrfr{DriftVecB}--\eqnrfr{ShadowLeg}) that
$$\eqalignno{%
\hbox to 3cm{$\displaystyle%
   {dx'^\mu_i\over dt} = {dx^\mu_i\over dt}$\hfil}
      &\hskip2cm {\rm for\ }t\ge t_0\cr
\hbox to 3cm{$\displaystyle%
   {dL'_{ij}\over dt} = {dL_{ij}\over dt}$\hfil}
      &\hskip2cm {\rm for\ }t\ge t_0\cr
\noalign{\rm while}
\hbox to 3cm{$\displaystyle%
   {d\alpha'_{ij}\over dt} = {d\alpha_{ij}\over dt} = 0$\hfil}%
     &\hskip2cm {\rm for\ }t = t_0\cr}
$$
Using the above we find that equations
(\eqnrfr{KmunuLegs}--\eqnrfr{KmunuAngles}) may be reduced to
$$
\openup3pt
\eqalignno{%
L_{ij}{dL_{ij}\over dt} &= -K_{\mu\nu}\Delta x^\mu_{ij}\Delta x^\nu_{ij}
&(\eqnrfr{KmunuLegs}')\cr
0 &= -2K_{\mu\nu}\Delta x^\mu_{oi} x^\nu_{oj}
&(\eqnrfr{KmunuAngles}')\cr}
$$
on $t=t_0$. From $(\eqnrfr{KmunuAngles}')$ we see immediately that $K_{\mu\nu}=0$
for $\mu\not=\nu$. The remaining equations are of the form
$$
\Displaylines{%
L_{oi^\pm}{dL_{oi^\pm}\over dt}  = - K_{ii} L^2_{oi^\pm}\cr
L_{i^\pm j^\pm}{dL_{i^\pm j^\pm }\over dt} = - K_{ii} L^2_{oi^\pm}
                                             - K_{jj} L^2_{oj^\pm}\cr}
$$
for $i\not =j$ drawn from the set $\lbrace 1,2,3\rbrace$. We can see that the second equation
is a trivial consequence of the first equation and the assumed constraint on the
diagonals. Thus this second equation is redundant and may be excluded when
solving for the $K_{\mu\nu}$ on $t=t_0$. Thus we obtain the following
equations for the $K_{\mu\nu}$
$$
\Displaylines{%
L_{01^\p}{dL_{01^\p}\over dt}  = - K_{11} L^2_{01^\p}\hskip 2cm
L_{01^\m}{dL_{01^\m}\over dt}  = - K_{11} L^2_{01^\m}\cr
L_{02^\p}{dL_{02^\p}\over dt}  = - K_{22} L^2_{02^\p}\hskip 2cm
L_{02^\m}{dL_{02^\m}\over dt}  = - K_{22} L^2_{02^\m}\cr
L_{03^\p}{dL_{03^\p}\over dt}  = - K_{33} L^2_{03^\p}\hskip 2cm
L_{03^\m}{dL_{03^\m}\over dt}  = - K_{33} L^2_{03^\m}\cr}
$$
This is clearly an overdetermined system for $K_{11}, K_{22}$ and $K_{33}$.
One could choose to take an average solution, such as,
$$
\openup5pt
\eqalign{%
K_{11} &= {-1\over2}\left({1\over L_{01^\p}}{dL_{01^\p}\over dt}
                         +{1\over L_{01^\m}}{dL_{01^\m}\over dt}\right)\cr
K_{22} &= {-1\over2}\left({1\over L_{02^\p}}{dL_{02^\p}\over dt}
                         +{1\over L_{02^\m}}{dL_{02^\m}\over dt}\right)\cr
K_{33} &= {-1\over2}\left({1\over L_{03^\p}}{dL_{03^\p}\over dt}
                         +{1\over L_{03^\m}}{dL_{03^\m}\over dt}\right)\cr}
$$
(which we would also get from a least square or singular value decomposition).
However, we have yet to impose precisely the condition that the lattice be
homogeneous. 

Consider two neighbouring vertices $p$ and ${\Bar p}$ and their respective RNC
cells. Let $x^\mu$ and ${\Bar x}^\mu$ be the RNC coordinates for the vertices
$p$ and ${\Bar p}$ respectively. We can align the coordinates such that the two
vertices lie on, say, the $x$-axis. We can now impose homogeneity by demanding
that $g_{\mu\nu}$ and $K_{\mu\nu}$ of the RNC for ${\Bar p}$ be obtained by Lie
dragging the $g_{\mu\nu}$ and $K_{\mu\nu}$ of the RNC for $p$ along the integral
curves of $\partial/\partial x$. For our lattice the 3-metric is flat everywhere
so this condition reduces to just a statement about the extrinsic curvatures,
$$
K_{\mu\nu}(p) = {\Bar K}_{\mu\nu}({\Bar p})
$$
This now forces us to make an alternative choice in solving the above
system for $K_{\mu\nu}$, namely,
$$
\eqalignno{%
K_{11} &= {-1\over L_{01^\p}}{dL_{01^\p}\over dt}
        = {-1\over L_{01^\m}}{dL_{01^\m}\over dt}&\eqndef{KmunuA}\cr
K_{22} &= {-1\over L_{02^\p}}{dL_{02^\p}\over dt}
        = {-1\over L_{02^\m}}{dL_{02^\m}\over dt}&\eqndef{KmunuB}\cr
K_{33} &= {-1\over L_{03^\p}}{dL_{03^\p}\over dt}
        = {-1\over L_{03^\m}}{dL_{03^\m}\over dt}&\eqndef{KmunuC}\cr}
$$
The initial data must be chosen to obey these constraints.

A further consequence of this definition of homogeneity is that
$$
0=K_{\mu\nu,\alpha}=K_{\mu\nu\vert\alpha}
$$
the second equality following from the RNC condition that
$\Gamma^\mu_{\alpha\beta}=0$ at the origin. Thus we see that the momentum
constraint is identically satisfied. Since $K_{\mu\nu}$ is diagonal we also 
find that the Hamiltonian constraint (\eqnrfr{HamltnConst}) is just
$$
0 = K_{11}K_{22} + K_{11}K_{33} + K_{22}K_{33}
\eqno\eqndef{FinalHamltnConst}
$$

The main evolution equations (\eqnrfr{DkDt}) are now
$$\openup8pt
\Displaylines{%
{dK_{11}\over dt} = (-K_{11} + K_{22} + K_{33})K_{11}
&\eqndef{DKmunuA}\cr
{dK_{22}\over dt} = (+K_{11} - K_{22} + K_{33})K_{22}
&\eqndef{DKmunuB}\cr
{dK_{33}\over dt} = (+K_{11} + K_{22} - K_{33})K_{33}
&\eqndef{DKmunuC}\cr}
$$
while
$$
{dK_{\mu\nu}\over dt} = 0
\eqno\eqndef{ZeroDKmunu}
$$
for $\mu\not = \nu$.

We can see that these equations, when coupled with
(\eqnrfr{KmunuA}--\eqnrfr{KmunuC}), provide second order evolution equations
for the $L_{oi}$. We also need second order evolution equations for
$\alpha_{ij}$ and $L_{ij}$.

Using $K_{\mu\nu}=0, dK_{\mu\nu}/dt=0$ on $t=t_0$ for $\mu\not=\nu$ we find, by
differentiation of $(\eqnrfr{KmunuAngles}')$, that
$$
{d^2\alpha_{ij}\over dt^2} = 0
\eqno\eqndef{DDAlpha}
$$
on $t=t_0$, while differentiation of $(\eqnrfr{KmunuLegs}')$ leads to
$$
{d\over dt}\left(L_{ij}{dL_{ij}\over dt}\right) =
{d\over dt}\left(L_{oi}{dL_{oi}\over dt}+L_{oj}{dL_{oj}\over dt}\right)
\eqno\eqndef{DDLij}
$$
also on $t=t_0$. These are the required evolution equations for $\alpha_{ij}$
and $L_{ij}$. These equations are nothing more than the second
time derivatives of the original constraints on the angles and diagonals,
namely, $\alpha_{ij}=\pi/2$ and $L^2_{ij} = L^2_{oi} + L^2_{oj}$.

We are now in a position to fully analyse the future evolution of the
lattice. For our lattice the initial data are $L_{oi},L_{ij},\alpha_{ij}$ and
their first time derivatives. The complete set of evolution equations are
(\eqnrfr{KmunuA}--\eqnrfr{KmunuC},\eqnrfr{DKmunuA}--\eqnrfr{DDLij}).
There is only one initial value constraint, equation (\eqnrfr{FinalHamltnConst}).

From equations (\eqnrfr{DDAlpha}) and (\eqnrfr{DDLij}) we see that our initial
constraints on $\alpha_{ij}$ and the diagonals $L_{ij}$ are preserved by the
evolution equations. Equation (\eqnrfr{ZeroDKmunu}) shows that $K_{\mu\nu}$
remain diagonal. The $dL_{oi}/dt$ are updated from equations
(\eqnrfr{KmunuA}--\eqnrfr{KmunuC}) and thus the lattice remains homogeneous.

In summary, the smooth lattice equations guarantee that the future 3-dimensional
lattices remain flat and homogeneous.

By analogy with the Kasner form of the metric we might seek a solution of
the form
$$
\eqalign{%
L_{01^\p} &= L_{01^\m} = A t^{p_x}\cr
L_{02^\p} &= L_{02^\m} = B t^{p_y}\cr
L_{03^\p} &= L_{03^\m} = C t^{p_z}\cr}
$$
where $A,B,C$ are constants. It can seen that this is a solution of the
constraint (\eqnrfr{FinalHamltnConst}) and the evolution equations
(\eqnrfr{DKmunuA}--\eqnrfr{DKmunuC}) provided
$$
1 = p_x + p_y + p_z = p^2_x + p^2_y + p^2_z
$$
in exact agreement with the Kasner metric.

We conclude that for this simple model there is no discretisation error. It
should be noted that the same observation would occur if one had employed a
finite difference approach. This is a rather trivial example as it does not
touch upon some of the more delicate aspects of the method. The extrinsic
curvatures where computed without resort to least squares and the assumption of
homogeneity allowed us to set $N_{\vert\mu\nu}$ and $K_{\mu\nu\vert\alpha}$ to
zero. There was also no need to interpolate the future values of $K_{\mu\nu}$
back to the updated lattice. In a more general setting each of these aspects can
be expected to require some attention.

\beginsection{Discussion}

There have been other attempts to adapt the ADM 3+1 equations to a lattice. In
particular there are the works of Piran and Williams \Ref(4) and Friedmann and
Jack \Ref(5). In both cases the equations were presented in the context of the
Regge Calculus. This is another lattice method in which the metric is assumed to
be flat inside each pair of adjacent tetrahedra. The principle difference between
their approach and ours is in the way in which the ADM equations were imposed on
the lattice. In Piran+Williams and Friedmann+Jack the equations of motion for the
lattice were obtained from the ADM 3+1 action principle. They were unable to
evaluate the action directly because in the Regge Calculus one does not have
direct access to quantities such as $R_{\mu\nu}, N_{\vert\mu\nu}$ etc. Instead
one has pure scalar quantities such as the defect angle, the areas and volumes of
simplices. Thus in developing their equations both pairs of authors needed to
make many (reasonable) assumptions about how various terms in the standard ADM
action could be translated into a Regge form. This is a somewhat adhoc method
and thus casts some doubt on the validity of the final equations. Indeed the
equations given by Friedmann and Jack differ from those given by Piran and
Williams (though they do agree for appropriate choices of lapse and shift).

In contrast our equations have been derived in a systematic manner. The use of
Riemann normal coordinates has allowed us to extract in a very natural way all
of the relevant coordinate data from the lattice data. This in turn has lead to
a very natural adaptation of the ADM 3+1 equations to the lattice. Despite this
there are still many variations that need to be explored. Some of these are of a
minor numerical nature (eg.\ should all the $dL_{ij}/dt$ be used in estimating
$K_{\mu\nu}$ or just a limited set?). Others are potentially much more important,
in particular does the use of overdetermined systems lead to errors and
instabilities in the evolution of the lattice? There is also the major question
of how one can solve the initial value constraint equations. Despite these
concerns, we believe that the basics of our algorithm are well founded and
should survive in any subsequent lattice method for numerical relativity.

However, the proof of the pudding is always in the eating. The method has been
shown to have no discretisation error for the rather trivial example of the
Kasner cosmology. This is encouraging but it must be noted that this example
avoided many of the potential problems associated with this method. So it is
imperative that the proposed method be put to a serious test. We have already
successfully constructed the initial data for a time symmetric initial data
slice in the Schwarzschild spacetime \Ref(1). We are currently applying our
proposed method to evolve that data. The results will be reported (soon) in a
later paper.

%
%

\bgroup
\beginsection{References}
\parskip=0pt plus 0pt minus 4pt
\vskip 0.5cm

\R 1!Brewin, L.C.!
     Riemann normal coordinates, smooth lattices and numerical relativity.!
     Applied Mathematics Preprint. 97/3 also available at {\tt gr-qc/9701057}
     and\hfil\break
     {\tt http://newton.maths.monash.edu.au:8000/preprints/slgr.ps.gz}\par

\R 2!Brewin, L.C.!
     Truncation errors in estimating the curvatures of smooth lattices.!
     In preparation.\par

\R 3!Anninos, P. Daues, G., Masso, J., Seidel, E. and Wai-Mo Suen.!
     Horizon boundary condition for black hole spacetimes.!
     Phys.Rev.D. Vol.51(1995) pp.5562-5578.\par

\R 4!Piran, T. and Williams, R.M.!
     Three-plus-one formulation of Regge calculus.!
     Phys.Rev.D. Vol.33(1986) pp.1622-1633.\par

\R 5!Friedman, J.L. and Jack, I.!
     3+1 Regge calculus with conserved momentum and Hamiltonian constraints.!
     J.Math.Phys. Vol.27(1986) pp.2973-2986.\par

\egroup
     
\vfill\eject


%
%
%
\newdimen\rotdimen
\def\vspec#1{\special{ps:#1}}
\def\rotstart#1{\vspec{gsave currentpoint currentpoint translate
   #1 neg exch neg exch translate}}
\def\rotfinish{\vspec{currentpoint grestore moveto}}
%
%
\def\rotr#1{\rotdimen=\ht#1\advance\rotdimen by\dp#1%
   \hbox to\rotdimen{\hskip\ht#1\vbox to\wd#1{\rotstart{90 rotate}%
   \box#1\vss}\hss}\rotfinish}
%
%
\def\rotl#1{\rotdimen=\ht#1\advance\rotdimen by\dp#1%
   \hbox to\rotdimen{\vbox to\wd#1{\vskip\wd#1\rotstart{270 rotate}%
   \box#1\vss}\hss}\rotfinish}%
%
%
\def\rotu#1{\rotdimen=\ht#1\advance\rotdimen by\dp#1%
   \hbox to\wd#1{\hskip\wd#1\vbox to\rotdimen{\vskip\rotdimen
   \rotstart{-1 dup scale}\box#1\vss}\hss}\rotfinish}%
%
%
\def\rotf#1{\hbox to\wd#1{\hskip\wd#1\rotstart{-1 1 scale}%
   \box#1\hss}\rotfinish}%
%
%

 \ifx\MYUNDEFINED\BoxedEPSF
   \let\temp\relax
 \else
   \message{}
   \message{ !!! BoxedEPS %
         or BoxedArt macros already defined !!!}
   \let\temp 
 \fi
  \temp
 
 \chardef\CatAt\the\catcode`\@
 \catcode`\@=11
 \chardef\C@tColon\the\catcode`\:
 \chardef\C@tSemicolon\the\catcode`\;
 \chardef\C@tQmark\the\catcode`\?
 \chardef\C@tEmark\the\catcode`\!

 \def\PunctOther@{\catcode`\:=12
   \catcode`\;=12 \catcode`\?=12 \catcode`\!=12}
 \PunctOther@

 \let\wlog@ld\wlog 
 \def\wlog#1{\relax} 

 \newif\ifIN@
 \newdimen\XShift@ \newdimen\YShift@ 
 \newtoks\Realtoks
 
  %
 \newdimen\Wd@ \newdimen\Ht@
 \newdimen\Wd@@ \newdimen\Ht@@
 \newdimen\TT@
 \newdimen\LT@
 \newdimen\BT@
 \newdimen\RT@
 \newdimen\XSlide@ \newdimen\YSlide@ 
 \newdimen\TheScale  
 \newdimen\FigScale  
 \newdimen\ForcedDim@@

 \newtoks\EPSFDirectorytoks@
 \newtoks\EPSFNametoks@
 \newtoks\BdBoxtoks@
 \newtoks\LLXtoks@  
 \newtoks\LLYtoks@

 \newif\ifNotIn@
 \newif\ifForcedDim@
 \newif\ifForceOn@
 \newif\ifForcedHeight@
 \newif\ifPSOrigin

 \newread\EPSFile@ 
 
  \def\ms@g{\immediate\write16}

 \newif\ifIN@\def\IN@{\expandafter\INN@\expandafter}
  \long\def\INN@0#1@#2@{\long\def\NI@##1#1##2##3\ENDNI@
    {\ifx\m@rker##2\IN@false\else\IN@true\fi}%
     \expandafter\NI@#2@@#1\m@rker\ENDNI@}
  \def\m@rker{\m@@rker}

  \newtoks\Initialtoks@  \newtoks\Terminaltoks@
  \def\SPLIT@{\expandafter\SPLITT@\expandafter}
  \def\SPLITT@0#1@#2@{\def\TTILPS@##1#1##2@{%
     \Initialtoks@{##1}\Terminaltoks@{##2}}\expandafter\TTILPS@#2@}


  \newtoks\Trimtoks@

 \def\ForeTrim@{\expandafter\ForeTrim@@\expandafter}
 \def\ForePrim@0 #1@{\Trimtoks@{#1}}
 \def\ForeTrim@@0#1@{\IN@0\m@rker. @\m@rker.#1@%
     \ifIN@\ForePrim@0#1@%
     \else\Trimtoks@\expandafter{#1}\fi}

  \def\Trim@0#1@{%
      \ForeTrim@0#1@%
      \IN@0 @\the\Trimtoks@ @%
        \ifIN@ 
             \SPLIT@0 @\the\Trimtoks@ @\Trimtoks@\Initialtoks@
             \IN@0\the\Terminaltoks@ @ @%
                 \ifIN@
                 \else \Trimtoks@ {FigNameWithSpace}%
                 \fi
        \fi
      }


   \newtoks\pt@ks
   \def \getpt@ks 0.0#1@{\pt@ks{#1}}
   \dimen0=0pt\expandafter\getpt@ks\the\dimen0@

  \newtoks\Realtoks
  \def\Real#1{%
    \dimen2=#1%
      \SPLIT@0\the\pt@ks @\the\dimen2@
       \Realtoks=\Initialtoks@
            }

   \newdimen\Product
   \def\Mult#1#2{%
     \dimen4=#1\relax
     \dimen6=#2%
     \Real{\dimen4}%
     \Product=\the\Realtoks\dimen6%
        }

 \newdimen\Inverse
 \newdimen\hmxdim@ \hmxdim@=8192pt
 \def\Invert#1{%
  \Inverse=\hmxdim@
  \dimen0=#1%
  \divide\Inverse \dimen0%
  \multiply\Inverse 8}

   \def\Rescale#1#2#3{
              \divide #1 by 100\relax
              \dimen2=#3\divide\dimen2 by 100 \Invert{\dimen2}%
              \Mult{#1}{#2}%
              \Mult\Product\Inverse 
              #1=\Product}

  \def\Scale#1{\dimen0=\TheScale %
      \divide #1 by  1280 
      \divide \dimen0 by 5120 %
      \multiply#1 by \dimen0 
      \divide#1 by 10   
     }
 

 \newbox\scrunchbox

 \def\Scrunched#1{{\setbox\scrunchbox\hbox{#1}%
   \wd\scrunchbox=0pt
   \ht\scrunchbox=0pt
   \dp\scrunchbox=0pt
   \box\scrunchbox}}

 \def\Shifted@#1{%
   \vbox {\kern-\YShift@
       \hbox {\kern\XShift@\hbox{#1}\kern-\XShift@}%
           \kern\YShift@}}


 \def\cBoxedEPSF#1{{}\leavevmode 
   \ReadNameAndScale@{#1}%
   \SetEPSFSpec@
   \ReadEPSFile@ \ReadBdB@x  
     \TrimFigDims@ 
     \CalculateFigScale@  
     \ScaleFigDims@
     \SetInkShift@
   \hbox{$\mathsurround=0pt\relax
         \vcenter{\hbox{%
             \FrameSpider{\hskip-.4pt\vrule}%
             \vbox to \Ht@{\offinterlineskip\parindent=\z@%
                \FrameSpider{\vskip-.4pt\hrule}\vfil 
                \hbox to \Wd@{\hfil}%
                \vfil
                \InkShift@{\EPSFSpecial{\EPSFSpec@}{\FigSc@leReal}}%
             \FrameSpider{\hrule\vskip-.4pt}}%
         \FrameSpider{\vrule\hskip-.4pt}}}%
     $}%
    \CleanRegisters@ 
    \ms@g{ *** Box composed for the %
         EPSF file \the\EPSFNametoks@}%
    }
 
 \def\tBoxedEPSF#1{\setbox4\hbox{\cBoxedEPSF{#1}}%
     \setbox4\hbox{\raise -\ht4 \hbox{\box4}}%
     \box4
      }

 \def\bBoxedEPSF#1{\setbox4\hbox{\cBoxedEPSF{#1}}%
     \setbox4\hbox{\raise \dp4 \hbox{\box4}}%
     \box4
      }

  \let\BoxedEPSF\cBoxedEPSF

   %

   %
  \def\gLinefigure[#1scaled#2]_#3{%
        \BoxedEPSF{#3 scaled #2}}
    
   %

  \def\EPSFxsize{\afterassignment\ForceW@\ForcedDim@@}
      \def\ForceW@{\ForcedDim@true\ForcedHeight@false}
  
  \def\EPSFysize{\afterassignment\ForceH@\ForcedDim@@}
      \def\ForceH@{\ForcedDim@true\ForcedHeight@true}

  %
 \def\ReadNameAndScale@#1{\IN@0 scaled@#1@
   \ifIN@\ReadNameAndScale@@0#1@%
   \else \ReadNameAndScale@@0#1 scaled\DefaultMilScale @
   \fi}
  
 \def\ReadNameAndScale@@0#1scaled#2@{
    \let\OldBackslash@\\%
    \def\\{\OtherB@ckslash}%
    \edef\temp@{#1}%
    \Trim@0\temp@ @%
    \EPSFNametoks@\expandafter{\the\Trimtoks@ }%
    \FigScale=#2 pt%
    \let\\\OldBackslash@
    }
 
 \def\SetDefaultEPSFScale#1{%
      \global\def\DefaultMilScale{#1}}

 \SetDefaultEPSFScale{1000}

  %
 \def \SetBogusBbox@{%
     \global\BdBoxtoks@{ BoundingBox:0 0 100 100 }%
     \global\def\BdBoxLine@{ BoundingBox:0 0 100 100 }%
     \ms@g{ !!! Will use placeholder !!!}%
     }

 \def\ReadEPSFile@{
     \openin\EPSFile@\EPSFSpec@
     \relax  
  \ifeof\EPSFile@
     \ms@g{}%
     \ms@g{ !!! EPS FILE \the\EPSFDirectorytoks@
       \the\EPSFNametoks@\ WAS NOT FOUND !!!}
     \SetBogusBbox@
  \else
   \begingroup
   \catcode`\%=12\catcode`\:=12\catcode`\!=12
   \catcode`\G=14\catcode`\\=14\relax
   \global\read\EPSFile@ to \BdBoxLine@
   \IN@0!PS@\BdBoxLine@ @%
   \ifIN@
     \NotIn@true 
     \loop   
       \ifeof\EPSFile@\NotIn@false 
         \ms@g{}%
         \ms@g{ !!! BoundingBox NOT FOUND IN %
            \the\EPSFDirectorytoks@\the\EPSFNametoks@\ !!! }%
         \SetBogusBbox@
       \else\global\read\EPSFile@ to \BdBoxLine@
       \fi
       \global\BdBoxtoks@\expandafter{\BdBoxLine@}%
       \IN@0BoundingBox:@\the\BdBoxtoks@ @%
       \ifIN@\NotIn@false\fi%
     \ifNotIn@\repeat
   \else
         \ms@g{}%
         \ms@g{ !!! \the\EPSFNametoks@\ not PS!\  !!!}%
         \SetBogusBbox@
   \fi
  \endgroup\relax
  \fi
  \closein\EPSFile@ 
   }

  \def\ReadBdB@x{
   \expandafter\ReadBdB@x@\the\BdBoxtoks@ @}
  
  \def\ReadBdB@x@#1BoundingBox:#2@{
    \ForeTrim@0#2@%
    \IN@0atend@\the\Trimtoks@ @
       \ifIN@\Trimtoks@={0 0 100 100 }
         \ms@g{}%
         \ms@g{ !!! BoundingBox not found in %
         \the\EPSFDirectorytoks@\the\EPSFNametoks@\space !!!}%
         \ms@g{ !!! It must not be at end of EPSF !!!}%
         \ms@g{ !!! Will use placeholder !!!}%
       \fi
    \expandafter\ReadBdB@x@@\the\Trimtoks@ @%
   }
    
  \def\ReadBdB@x@@#1 #2 #3 #4@{
      \Wd@=#3bp\advance\Wd@ by -#1bp%
      \Ht@=#4bp\advance\Ht@ by-#2bp%
       \Wd@@=\Wd@ \Ht@@=\Ht@ 
       \LLXtoks@={#1}\LLYtoks@={#2}
      \ifPSOrigin\XShift@=-#1bp\YShift@=-#2bp\fi 
     }

   %
   \def\G@bbl@#1{}
   \bgroup
     \global\edef\OtherB@ckslash{\expandafter\G@bbl@\string\\}
   \egroup

  \def\SetEPSFDirectory{
           \bgroup\PunctOther@\relax
           \let\\\OtherB@ckslash
           \SetEPSFDirectory@}

 \def\SetEPSFDirectory@#1{
    \edef\temp@{#1}%
    \Trim@0\temp@ @
    \global\toks1\expandafter{\the\Trimtoks@ }\relax
    \egroup
    \EPSFDirectorytoks@=\toks1
    }

 \def\SetEPSFSpec@{%
     \bgroup
     \let\\=\OtherB@ckslash
     \global\edef\EPSFSpec@{%
        \the\EPSFDirectorytoks@\the\EPSFNametoks@}%
     \global\edef\EPSFSpec@{\EPSFSpec@}%
     \egroup}

  %
 \def\TrimTop#1{\advance\TT@ by #1}
 \def\TrimLeft#1{\advance\LT@ by #1}
 \def\TrimBottom#1{\advance\BT@ by #1}
 \def\TrimRight#1{\advance\RT@ by #1}

 \def\TrimFigDims@{%
    \advance\Wd@ by -\LT@ 
    \advance\Wd@ by -\RT@ \RT@=\z@
    \advance\Ht@ by -\TT@ \TT@=\z@
    \advance\Ht@ by -\BT@ 
    }

  %
  \def\ForceWidth#1{\ForcedDim@true
       \ForcedDim@@#1\ForcedHeight@false}
  
  \def\ForceHeight#1{\ForcedDim@true
       \ForcedDim@@=#1\ForcedHeight@true}

  \def\ForceOn{\ForceOn@true}
  \def\ForceOff{\ForceOn@false\ForcedDim@false}
  
  \def\epsfxsize{\afterassignment\ForceW@\ForcedDim@@}
      \def\ForceW@{\ForcedDim@true\ForcedHeight@false}
  
  \def\epsfysize{\afterassignment\ForceH@\ForcedDim@@}
      \def\ForceH@{\ForcedDim@true\ForcedHeight@true}
  
  \def\CalculateFigScale@{%
     \ifForcedDim@\FigScale=1000pt
           \ifForcedHeight@
                \Rescale\FigScale\ForcedDim@@\Ht@
           \else
                \Rescale\FigScale\ForcedDim@@\Wd@
           \fi
     \fi
     \Real{\FigScale}%
     \edef\FigSc@leReal{\the\Realtoks}%
     }
   
  \def\ScaleFigDims@{\TheScale=\FigScale
      \ifForcedDim@
           \ifForcedHeight@ \Ht@=\ForcedDim@@  \Scale\Wd@
           \else \Wd@=\ForcedDim@@ \Scale\Ht@
           \fi
      \else \Scale\Wd@\Scale\Ht@        
      \fi
      \ifForceOn@\relax\else\global\ForcedDim@false\fi
      \Scale\LT@\Scale\BT@  
      \Scale\XShift@\Scale\YShift@
      }
      

 \let\HideDisplacementBoxes\HideReservedBoxes  
 \let\ShowDisplacementBoxes\ShowReservedBoxes

  \ShowDisplacementBoxes
 
 \def\hSlide#1{\advance\XSlide@ by #1}
 \def\vSlide#1{\advance\YSlide@ by #1}
 
  \def\SetInkShift@{%
            \advance\XShift@ by -\LT@
            \advance\XShift@ by \XSlide@
            \advance\YShift@ by -\BT@
            \advance\YShift@ by -\YSlide@
             }
  \def\InkShift@#1{\Shifted@{\Scrunched{#1}}}
 
   %
  \def\CleanRegisters@{%
      \globaldefs=1\relax
        \XShift@=\z@\YShift@=\z@\XSlide@=\z@\YSlide@=\z@
        \TT@=\z@\LT@=\z@\BT@=\z@\RT@=\z@
      \globaldefs=0\relax}

 
 \def\SetTexturesEPSFSpecial{\PSOriginfalse
  \gdef\EPSFSpecial##1##2{\relax
    \edef\specialthis{##2}%
    \SPLIT@0.@\specialthis.@\relax
    \special{illustration ##1 scaled
                        \the\Initialtoks@}}}
 
  \def\SetUnixCoopEPSFSpecial{\PSOrigintrue 
   \gdef\EPSFSpecial##1##2{%
      \dimen4=##2pt
      \divide\dimen4 by 1000\relax
      \Real{\dimen4}
      \edef\Aux@{\the\Realtoks}%
      \includegraphics{##1\space}}}

  \def\SetBechtolsheimRokickiEPSFSpecial{\PSOrigintrue 
   \gdef\EPSFSpecial##1##2{%
      \dimen4=##2pt
      \divide\dimen4 by 1000\relax
      \Real{\dimen4}
      \edef\Aux@{\the\Realtoks}%
      \special{ps: psfiginit}%
      \special{ps: literal 1 1 0 0 1 1 startTexFig
           \the\mag\space 1000 div \Aux@\space mul 
           \the\mag\space 1000 div \Aux@\space mul scale}%
      \special{ps: include  ##1}%
      \special{ps: literal endTexFig}%
        }}

  \def\SetLisEPSFSpecial{\PSOrigintrue 
   \gdef\EPSFSpecial##1##2{%
      \dimen4=##2pt
      \divide\dimen4 by 1000\relax
      \Real{\dimen4}
      \edef\Aux@{\the\Realtoks}%
      \special{pstext="1 1 0 0 1 1 startTexFig\space
           \the\mag\space 1000 div \Aux@\space mul 
           \the\mag\space 1000 div \Aux@\space mul scale}%
      \includegraphics{##1}%
      \special{pstext=endTexFig}%
        }}

  \def\SetRokickiEPSFSpecial{\PSOrigintrue 
   \gdef\EPSFSpecial##1##2{%
      \dimen4=##2pt
      \divide\dimen4 by 10\relax
      \Real{\dimen4}
      \edef\Aux@{\the\Realtoks}%
      \includegraphics{##1}}}

  \def\SetInlineRokickiEPSFSpecial{\PSOrigintrue 
   \gdef\EPSFSpecial##1##2{%
      \dimen4=##2pt
      \divide\dimen4 by 1000\relax
      \Real{\dimen4}
      \edef\Aux@{\the\Realtoks}%
      \special{ps::[begin] 1 1 0 0 1 1 startTexFig\space
           \the\mag\space 1000 div \Aux@\space mul 
           \the\mag\space 1000 div \Aux@\space mul scale}%
      \special{ps: plotfile ##1}%
      \special{ps::[end] endTexFig}%
        }}

  \def\SetOzTeXEPSFSpecial{\PSOriginfalse 
  \gdef\EPSFSpecial##1##2{
     \special{##1\space 
       ##2 1000 div \the\mag\space 1000 div mul
       ##2 1000 div \the\mag\space 1000 div mul scale
       \the\LLXtoks@\space neg \the\LLYtoks@\space neg translate
             }}} 
 

 \def\SetArborEPSFSpecial{\PSOriginfalse 
   \gdef\EPSFSpecial##1##2{%
     \edef\specialthis{##2}%
     \SPLIT@0.@\specialthis.@\relax 
     \special{ps: epsfile ##1\space \the\Initialtoks@}}}

 \def\SetClarkEPSFSpecial{\PSOriginfalse 
   \gdef\EPSFSpecial##1##2{%
     \Rescale {\Wd@@}{##2pt}{1000pt}%
     \Rescale {\Ht@@}{##2pt}{1000pt}%
     \special{dvitops: import 
           ##1\space\the\Wd@@\space\the\Ht@@}}}



 \def\SetStandardEPSFSpecial{%
   \gdef\EPSFSpecial##1##2{%
     \ms@g{}
     \ms@g{%
       !!! Sorry! There is still no standard for \string%
       \special\ EPSF integration !!!}%
     \ms@g{%
      --- So you will have to identify your driver using a command}%
     \ms@g{%
      --- of the form \string\Set...EPSFSpecial, in order to get}%
     \ms@g{%
      --- your graphics to print.  See BoxedEPS.doc.}%
     \ms@g{}
     \KillEPSFSpecial
     }}

  \def\KillEPSFSpecial{\gdef\EPSFSpecial##1##2{}}

  \SetStandardEPSFSpecial 
 
 \let\wlog\wlog@ld 

 \catcode`\:=\C@tColon
 \catcode`\;=\C@tSemicolon
 \catcode`\?=\C@tQmark
 \catcode`\!=\C@tEmark

 \catcode`\@=\CatAt

 %
 %
 %
 %
 %

\HideDisplacementBoxes
\SetRokickiEPSFSpecial
\twelvepointsgl
\nopagenumbers
\overfullrule=0pt
\def\MyBig{\seventeenpointsgl}

\def\m{{\hbox{\tt-}}}
\def\p{{\hbox{\tt+}}}
\advance\baselineskip 2pt
\advance\vsize 1.0cm\relax
\lineskiplimit=0pt
\advance\baselineskip+1.5pt
%
\def\stomp#1{\setbox0=\hbox{#1}\dp0=0pt\ht0=0pt\box0}
%
%
\def\at(#1,#2)#3{\vbox to 0pt{\kern#2cm%
                 \hbox to 0pt{\kern#1cm\stomp{#3}\hss}\vss}\nointerlineskip}
%
\parindent=0pt
\newbox\boxa
\newbox\image
\newbox\xlabel
\newbox\ylabel
\newbox\zlabel
\newbox\caption
\newbox\imagea
\newbox\imageb
\newbox\xlabela
\newbox\ylabela
\newbox\zlabela
\newbox\xlabelb
\newbox\ylabelb
\newbox\zlabelb
\newbox\capta
\newbox\captb
%
\setbox\caption=\hbox to 15cm{\hsize=15cm\vtop{
{\bf Figure \figdef{DriftVec}}.\ This figures displays the drift vector
$\gamma^\mu_i$ and the relationships between the coordinates on the primary
and shadow lattices. Clearly
$\delta x^\mu_i = (-\gamma^\mu + N^\mu_i)\delta t$
leading directly to equation (\eqnrfr{DriftVecA}). We also have, by
construction, that $x^\mu_i=x'^\mu_i$ on $\Sigma_0$.
Thus $\delta x^\mu_i = \delta x'^\mu_i + \gamma^\mu_i \delta t$ which in
turn leads to equation (\eqnrfr{DriftVecB}).}}
\setbox\image=\hbox{\bBoxedEPSF{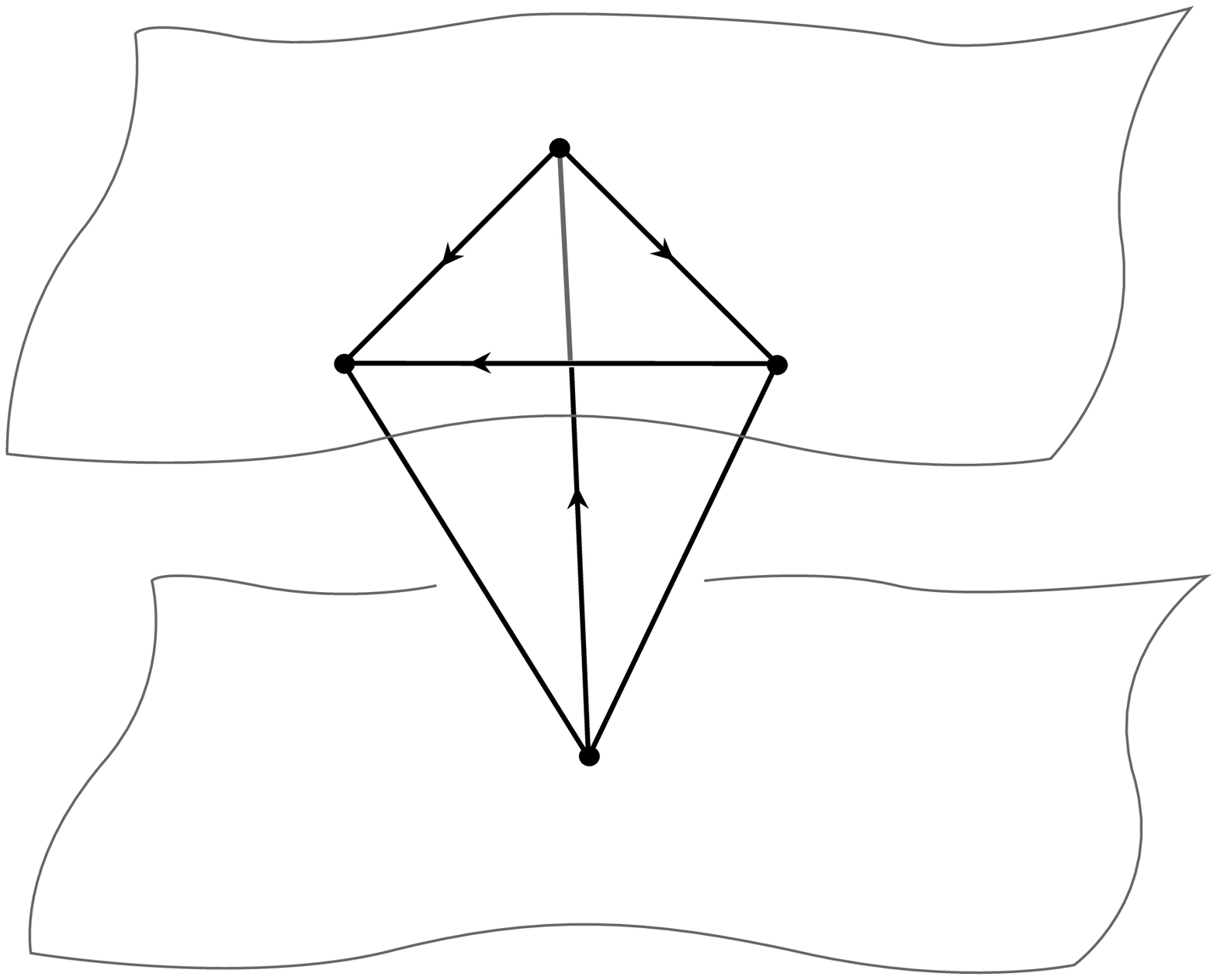 scaled 1000}}
\setbox\boxa=%
\hbox to 18.0cm{
\vtop to 18.5cm{
\MyBig
\at(  0.2,15.3){\box\image}
\at(  1.0,18.0){\box\caption}
\at( 10.0,12.1){$x^\mu_i=x'^\mu_i$}
\at(  9.3, 2.3){$x'^\mu_i+\delta x'^\mu_i$}
\at(  7.2, 8.0){$N{\Vec n}\delta t$}
\at(  7.9,12.1){$(i)$}
\at(  7.7, 2.3){$(i')$}
\at( 11.0, 4.0){$N^\mu_i\delta t$}
\at(  5.7, 4.0){$\gamma^\mu_i\delta t$}
\at( 12.5, 6.2){$x^\mu_i$}
\at(  2.8, 6.6){$x^\mu_i + \delta x^\mu_i$}
\at(  7.0, 5.5){$\delta x^\mu_i$}
\at(  4.5, 5.5){$(i)$}
\at(  3.5,10.5){$t=t_0$}
\at(  3.5, 2.0){$t=t_0+\delta t$}
\at( 16.0,10.5){$\Sigma_0$}
\at( 16.0, 2.0){$\Sigma_{\delta t}$}
\vfill}\hfill}
%
%
\centerline{\box\boxa}\vfill\eject
\setbox\caption=\hbox to 15cm{\hsize=15cm\vtop{
{\bf Figure \figdef{KmunuFitA}}.\ In the first pass over the lattice,
$K_{\mu\nu}$ is computed at each vertex ($\bullet$).
In a second pass, $K_{\mu\nu\vert\alpha}$ at the central vertex is
estimated by fitting the linear approximation
${\Tilde K}_{\mu\nu}(x) = {\Tilde K}_{\mu\nu} +
 {\Tilde K}_{\mu\nu\vert\alpha}x^\alpha$
to the data on each vertex. This will require a coordinate transformation from
neighbouring cells to get data on the boundary of the cell. The
$N_{\vert\mu\nu}$ at the central vertex can be estimated by first forming
estimates for $N_{\vert\mu}$ at the centre of each leg
(\vrule height 5pt width 5pt depth 0pt\relax). That data is then
approximated by a linear function
${\Tilde N}_{\vert\mu}(x) = {\Tilde N}_{\vert\mu} +
 {\Tilde N}_{\vert\mu\nu}x^\nu$. 
We can then use ${\Tilde N}_{\vert\mu\nu}$ as an estimate of 
$N_{\vert\mu\nu}$ at the central vertex.}}
\setbox\image=\hbox{\bBoxedEPSF{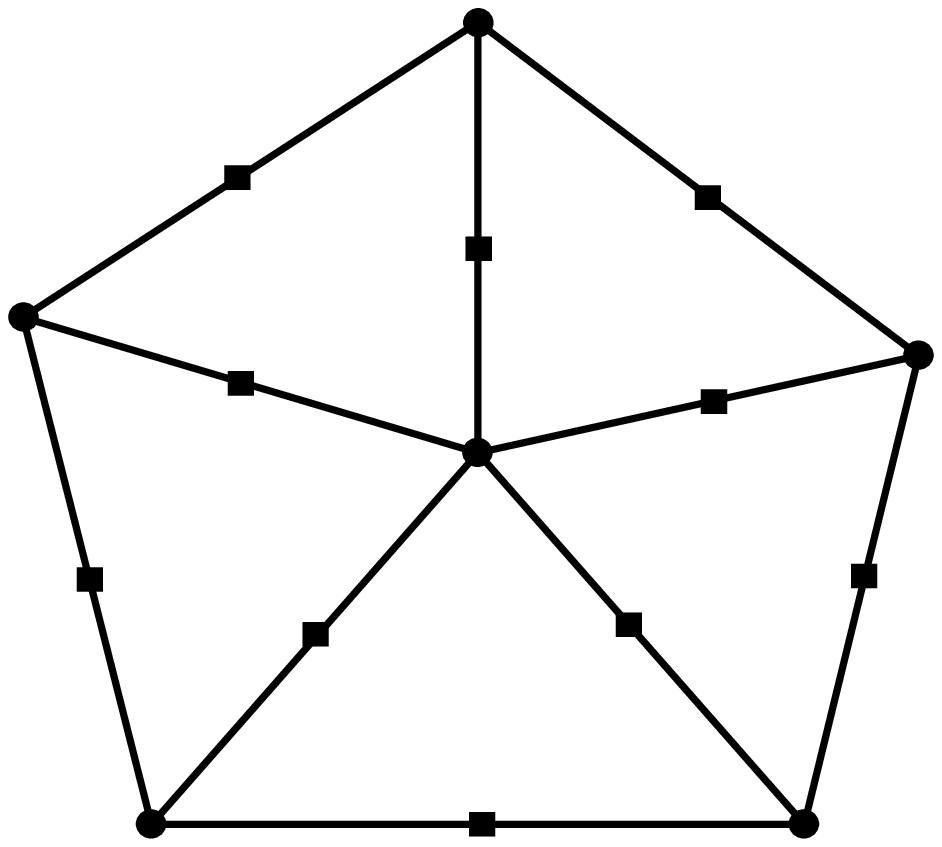 scaled 1000}}
\setbox\boxa=%
\hbox to 18.0cm{
\vtop to 18.5cm{
\at(  3.0,15.0){\box\image}
\at(  1.0,18.0){\box\caption}
\vfill}\hfill}
%
%
\centerline{\box\boxa}\vfill\eject
\setbox\caption=\hbox to 15cm{\hsize=15cm\vtop{
{\bf Figure \figdef{KmunuFitB}}.\ After one time step the shadow lattice has
drifted relative to the primary lattice. However the updated $K_{\mu\nu}$
are recorded on the shadow lattice. Thus an interpolation from the shadow
lattice back to the primary lattice is required. This can be done by forming
a linear approximation
$K_{\mu\nu}(x^\alpha) = {\Tilde K}_{\mu\nu} 
                      + {\Tilde K}_{\mu\nu\vert\alpha} x^\alpha$.
and evaluating it at $x^\alpha=0$. Note that no such interpolation is required
if the drift vectors are set to zero.}}
\setbox\image=\hbox{\bBoxedEPSF{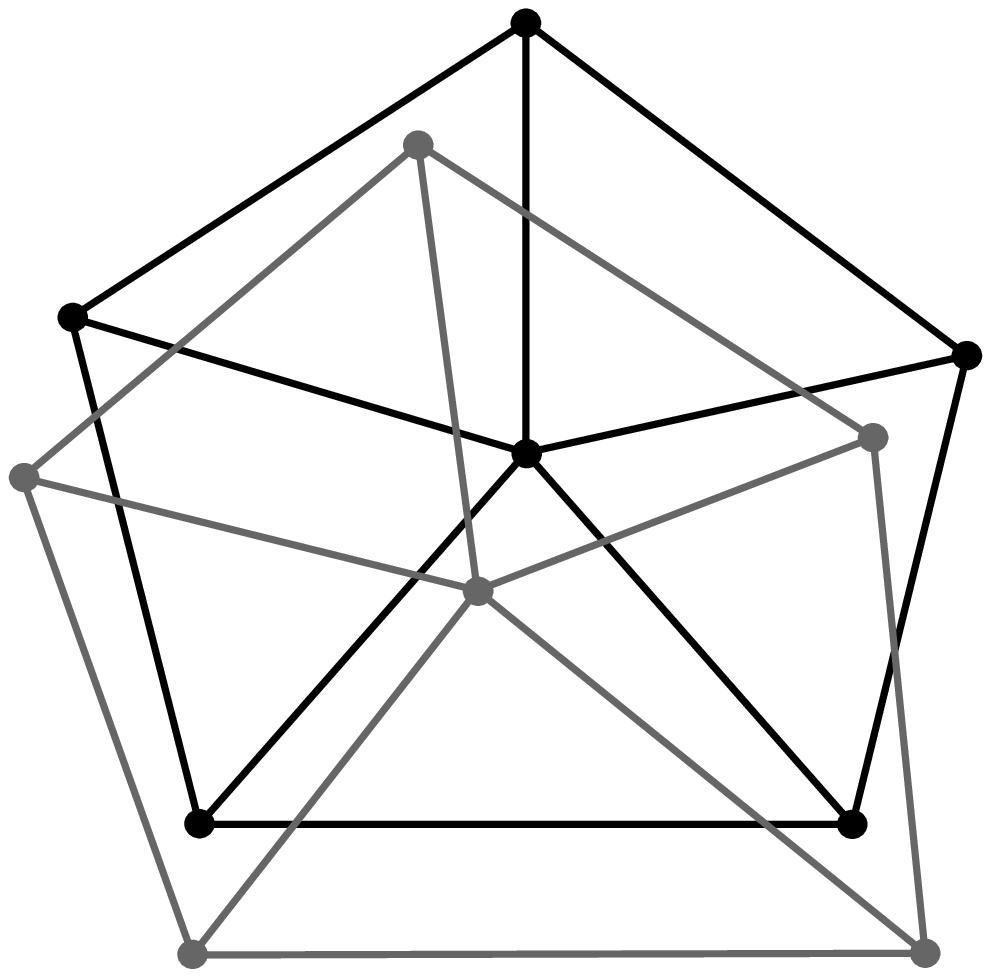 scaled 1000}}
\setbox\boxa=%
\hbox to 18.0cm{
\vtop to 18.5cm{
\at(  3.0,15.0){\box\image}
\at(  1.0,18.0){\box\caption}
\vfill}\hfill}
%
%
\centerline{\box\boxa}\vfill\eject
\setbox\caption=\hbox to 15cm{\hsize=15cm\vtop{%
{\bf Figure \figdef{DiamondCell}}.\ A typical computational cell in the cubic
lattice. This cell consists of eight tetrahedra attached to the central vertex.
The full cubic lattice can be obtained by replicating this structure throughout
the space. Note that this leads to overlapping tetrahedra. This typical cell
can be used in many spaces other than just the Kasner cosmology considered
in the text. The $T^3$ topology of the Kasner cosmology can be obtained by
identifying opposite ends of the cubic lattice. Note that, for clarity,
some of the legs have not be shown in this figure
(eg. $(1^\p 3^\m), (1^\m 2^\p)$ etc.)}}
\setbox\image=\hbox{\bBoxedEPSF{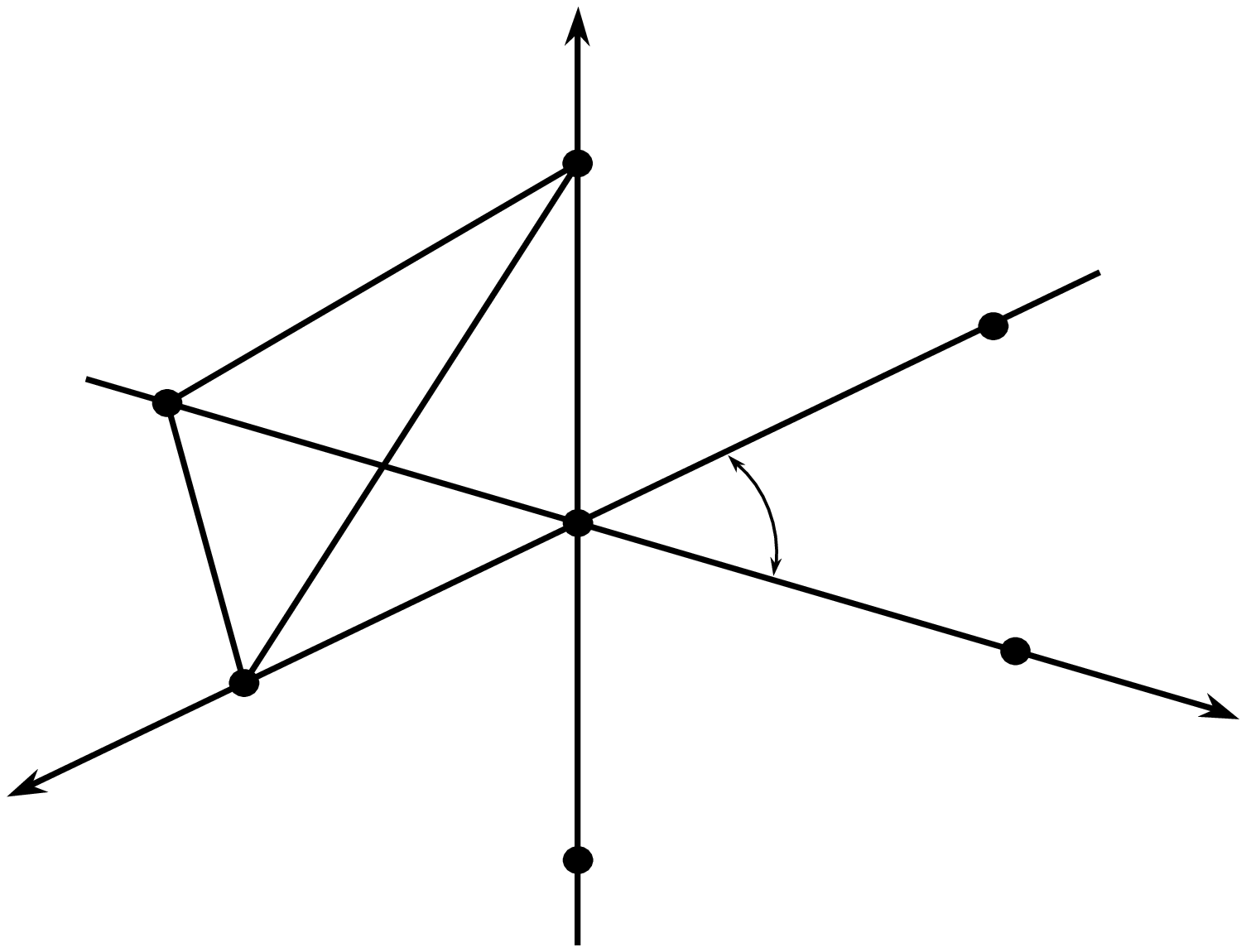 scaled 1000}}
\setbox\boxa=%
\hbox to 18.0cm{
\vtop to 18.5cm{
\MyBig
\at(  1.0,15.0){\box\image}
\at(  1.0,18.0){\box\caption}
\at(  8.7,10.8){$0$}
\at(  4.0,12.8){$1^\p$}
\at( 13.1, 8.4){$1^\m$}
\at( 12.8,12.4){$2^\p$}
\at(  1.9, 9.2){$2^\m$}
\at(  8.7, 5.5){$3^\p$}
\at(  8.7,14.2){$3^\m$}
\at( 10.9, 9.7){$\alpha_{1^\m 2^\p}$}
\at(  1.0,14.2){$X$}
\at( 16.0,13.2){$Y$}
\at(  8.7, 3.4){$Z$}
\vfill}\hfill}
%
%
\centerline{\box\boxa}\vfill\eject
\setbox\caption=\hbox to 15cm{\hsize=15cm\vtop{%
{\bf Figure \figdef{CubicLattice}}.\ An $xy$--cross-section of (part of) the
cubic lattice. The heavy solid lines denote the legs of one computational cell.
Notice that the diagonal legs attached to the central vertex are not included
in the computational cell. Note also how neighbouring cells overlap each other.}}
\setbox\image=\hbox{\bBoxedEPSF{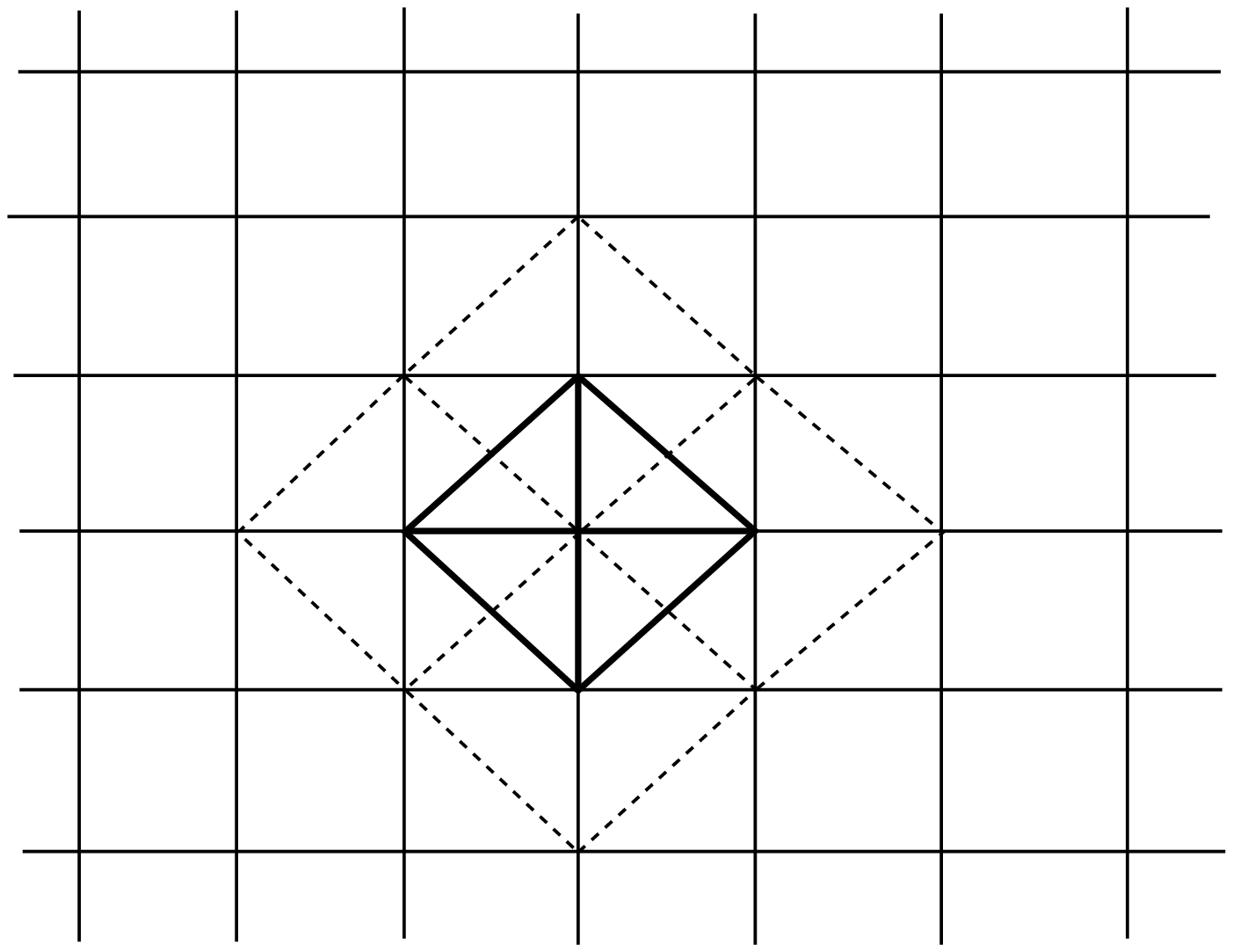 scaled 1000}}
\setbox\boxa=%
\hbox to 18.0cm{
\vtop to 18.5cm{
\at(  1.0,15.0){\box\image}
\at(  1.0,18.0){\box\caption}
\vfill}\hfill}
%
%
\centerline{\box\boxa}\vfill\eject

\bye